\documentclass[11pt]{article}

\newcounter{ctr}

\usepackage[dvips]{graphics}
\usepackage{epsfig}

\begin{document}

\begin{center}
{\Huge A generic mechanism for adaptive growth rate regulation}
\end{center}
\begin{large}
\begin{center}
Chikara Furusawa\footnotemark[1]\footnotemark[3]{} and 
Kunihiko Kaneko\footnotemark[2]\footnotemark[3]{}
\end{center}
\end{large}
\noindent
\footnotemark[1] Department of Bioinformatics Engineering, Graduate School of 
Information Science and Technology, Osaka University, 2-1 Yamadaoka, Suita, Osaka 565-0871, Japan\\
\footnotemark[2] Department of Pure and Applied Sciences, University of Tokyo,
Komaba, Meguro-ku, Tokyo 153-8902, Japan\\
\footnotemark[3] Complex Systems Biology Project, ERATO, JST, 
Komaba, Meguro-ku, Tokyo 153-8902, Japan\\

\noindent
{\bf Abstract}\\
\noindent
How can a microorganism adapt to a variety of environmental conditions
despite there exists a limited number of signal transduction
machineries?  
We show that for any growing cells whose gene expression
is under stochastic fluctuations, adaptive cellular state is inevitably
selected by noise, even without specific signal transduction network for it. 
In general, changes in protein concentration in a cell are
given by its synthesis minus dilution and degradation, both of which
are proportional to the rate of cell growth. In an adaptive state with
a higher growth speed, both terms are large and balanced.  
Under the presence of noise in gene expression, the
adaptive state is less affected by stochasticity since both the
synthesis and dilution terms are large, while for a non-adaptive state
both the terms are smaller so that cells are easily kicked out of the
original state by noise. 
Hence, escape time from a cellular state and the
cellular growth rate are negatively correlated. 
This leads to a selection of adaptive states with higher growth rates, and model
simulations confirm this selection to take place in general.  
The results suggest a general form of adaptation that has never been
brought to light - a process that requires no specific machineries for
sensory adaptation.  The present scheme may help explain a wide range
of cellular adaptive responses including the metabolic flux optimization
for maximal cell growth.

\section{Introduction}

Cells adapt to a variety of environmental conditions by changing the
pattern of gene expression and metabolic flux distribution. 
These adaptive responses are generally explained by signal transduction
mechanisms, where extracellular events are translated into
intracellular events through regulatory molecules. 
For example, the Lac operon of {\sl Escherichia coli} encodes proteins involved in lactose
metabolism, and expression of the operon is controlled by a regulatory
protein so that, when lactose is available, these proteins are
expressed in an efficient and coordinated manner \cite{Lac_operon}. 
In general, adaptive responses are depicted by a pre-wired logic circuit that
takes an environmental condition as an input and gene expression as an
output.

However, such program-like descriptions may not always apply, since
the number of possible environmental conditions to which a cell must
adapt is so large compared to the limited repertoire of gene
regulatory mechanisms. For example, experiments using phenotype
microarrays \cite{phenotype_microarray} revealed that when {\sl E. coli} cells grow in hundreds of
environmental conditions, including different carbon and nitrogen
sources and stress environments, in which they are distinctly altered states
of gene expression \cite{phenotype_microarra_Ecoli}. 
Considering that the number of {\sl E. coli}
genes categorized as 'signal transduction mechanisms' in the genome is
less than a few hundred \cite{COGs}, it is less plausible that cells have gene
regulatory programs to adapt to such a variety of environmental conditions\footnote[1]{
Of course, there could be a combinatorial ways in which regulation could
happen, as in seen in proteins with many interaction partners. In this case, however, selection of a
pathway for optimal growth might cause a combinatorially difficult problem.
In this sense, search of  an alternative mechanism other than the conventional
signal transduction, if any, will be relevant to understand cellular adaption.
}.
Indeed, in case of bacterial growth, a general adaptation
process that occurs over generations seems to exist in addition to
adaptation through gene regulation by signal transduction mechanisms
\cite{Egli,Ferenci}.

Recent two studies indicated the possibility that cells can respond to
environmental changes adaptively without pre-programmed signal
transduction mechanisms. Braun and colleagues demonstrated using yeast
cells that even when the promoter of the essential gene (HIS3) is
detached from the original regulatory system, expression of the
gene is regulated adaptively in response to environmental demands
\cite{Braun}. Furthermore, Kashiwagi et al. demonstrated that {\sl E. coli} cells
select an appropriate intra-cellular state according to environmental
conditions without the help of signal transduction machineries \cite{Kashiwagi-Yomo}.
There, an artificial gene network composed of two mutually
inhibitory operons was introduced into {\sl E. coli} cells, so that states of
gene expression are bistable. They found that the cells shift to the
adaptive cellular state by expressing the gene required to survive in
the environment.

In the present study, we demonstrate that cells select states most
favorable for their survival among a large number of other possible
states as an inevitable outcome of the very fact that cells grow and
that gene expression is inherently stochastic \cite{noise1, noise2, noise3, noise4, noise5}.  
By studying a model
that consists of a protein regulatory network and a metabolic reaction
network, we show that cellular states with high growth rates are
selected among a huge number of possible cellular states, only
mediated by fluctuations of gene expressions.  This
selection of a higher growth state is theoretically explained by
noting that a state with lower growth speed is more influenced by
stochasticity in gene expression, so that it is easily kicked away to
switch to a state with a higher growth rate. It is generally shown
that there is a negative correlation between the rate of noise-driven
escape from a given state and the cellular growth rate. Due to
this negative correlation, an optimal growth state is selected
spontaneously. Noting the generality of this selection mechanism, we
provide a possible answer to the question how cells generally adapt to
a huger variety of environmental conditions by changing their gene
expression pattern even without a specific signal transduction
mechanism.

\section{Results}

\subsection{Cell Model}

Fig.1 represents the schematic
representation of our model.  It consists of two networks,
i.e., a regulatory network which controls 
expression levels of proteins through each other, and a metabolic reaction network whose fluxes
are governed by expression levels of the proteins.  
The internal state of a cell is represented by a set of expression
levels of $n$ proteins $(x_1, x_2, \cdots, x_n)$ and concentrations of
$m$ metabolic substrates $(y_1, y_2, \cdots, y_m)$. The change in
expression levels of proteins over time is determined by (i) the
synthesis of proteins, (ii) dilution of proteins by the
cell volume growth, and (iii)molecular fluctuation arising from
stochasticity in chemical reactions\footnote[2]{One can also include the degradation process of
proteins, other than the dilution effect.  
Even if the degradation process is added to the model, 
we obtain qualitatively the same results.}.
The dilution of proteins is proportional to the growth rate of cell 
volume $v_g$, which is determined by the metabolic fluxes.
Also, we assume that the rates of protein synthesis
are proportional to the growth rate $v_g$.
This assumption is natural and is necessary to maintain a steady state, 
since the decrease in protein concentration by dilution due to the cell growth has to be
compensated by synthesis.
In terms of cell biology, it relied on the fact that expression of 
proteins and cellular growth basically share the common resources,
such as amino acids for their building blocks, and
ATP as energy source.  
Thus we write the dynamics of expression level
of the $i$-th protein as follows:
 
\begin{equation}
\frac{dx_i(t)}{dt} = f(\sum_{j=1}^n W_{ij}x_j(t) - \theta)v_g(t) - x_{i}(t)v_g(t) 
+ \eta(t)
\end{equation}

The first and second terms in r.h.s. represent synthesis,
dilution of the protein $i$, respectively. In the
first term, the regulation of protein expression levels by other
proteins are indicated by regulatory matrix $W_{ij}$, which takes 1, 0, or -1
representing activation, no regulatory interaction, and inhibition of
the $i$-th protein expression by the $j$-th protein, respectively.
The synthesis of proteins is given by the sigmoidal regulation function
$f(z)=1/(1+exp(-\mu z))$, where $z=(\sum{W_{ij}x_j(t)}-\theta)$ is the
total regulatory input with the threshold $\theta$ for activation of
synthesis, and $\mu$ indicates gain parameter of the sigmoid
function\footnote[3]{
See \cite{Mjolsness} and \cite{Sole} for this class of gene-network model.  
Also, instead of the above form one can use the form
$(z/\theta)^{n}/(1+(z/\theta)^{n})$ with Hill coefficient $n$.
}.
The regulatory interactions are determined randomly
with the connection rate $\rho_a$, $\rho_i$, indicating the connection
rate of excitatory paths and inhibitory paths, respectively.

The last term of r.h.s. in eq.(1) represent the molecular
fluctuation.  
For a specific form of the noise, we assume 
that there are fluctuations in the order of
$\sqrt{N}$ for reaction involving $N$ molecules, 
then we add a noise term $\eta=\xi(t) \sqrt{x_i(t)}$, where
$\xi(t)$ denotes Gaussian white noise satisfying $ <\xi(t) \xi(t')>= \sigma ^2
\delta(t-t')$, with $\sigma$ the amplitude of the noise.
In this model, we assume that 
the amplitude of the noise is independent of the 
synthesis and dilution terms of proteins, since 
the inclusion of synthesis and dilution dependent part of the noise
does not change the simulation results qualitatively.

Temporal change in concentrations of metabolic substrates are given by
metabolic reactions and transportation of substrates from the outside
of the cell\footnote[4]{
In the metabolic reaction dynamics, we neglect the effect
of fluctuations in the concentration of substrate, considering that
the numbers of substrate molecules in a cell are sufficiently large.
However, inclusion of noise term here does not introduce any essential
changes to the result to be discussed.}.
Each metabolic reaction is catalyzed by a corresponding protein.
Some nutrient substrates are
supplied from the environment by diffusion through the cell membrane,
to ensure the growth of a cell. Here, the dynamics of
$i$-th substrate concentration $y_i$ is represented as:

\begin{equation}
\frac{dy_i}{dt}=\epsilon \sum_{j=1}^n \sum_{k=1}^m Con(k,j,i)x_j y_k 
-\epsilon \sum_{j'=1}^n \sum_{k'=1}^m Con(i,j',k')x_{j'} y_i
+D(Y_i-y_i)
\end{equation}

\noindent
where $\epsilon$ indicates the coefficient for the metabolic
reactions, and $Con(i,j,k)$ represents the reaction matrix of the
metabolic network, which takes 1 if there is a metabolic reaction from
$i$-th substrate to $k$-th substrate catalyzed by $j$-th protein, and
0 otherwise. The first and second terms of r.h.s.
correspond to synthesis and consumption of $i$-th substrate by
metabolic reactions, respectively. The third term of r.h.s. represents
the transportation of the substrate through the cell membrane, which
is approximated by the linear term in the 
diffusion process with a diffusion coefficient
$D$. $Y_i$ is a constant representing the concentration of
$i$-th substrate in the environment. The concentration $Y_i$ is
nonzero only for nutrient substrates. 

The cellular growth rate $v_g$ is determined by the dynamics in the
metabolic reactions. We assume that some of metabolic substrates are
necessary for cellular growth, and the growth rate $v_g$ is determined
as a function of the concentrations of them.  Several choices of the function
are possible, and the results to be discussed are generally observed
as long as the growth rate varies drastically depending on the
concentrations. Here we assume that the growth rate is proportional to
the minimal concentration among these necessary substrates.  In other
words, among $m$ metabolic substrates there are $r$ substrates
$(y_1,y_2,\cdots,y_r)$ required for cellular growth, and the growth
rate is represented as $v_g \propto \min(y_1,y_2,\cdots,y_r)$.

In this study, we have studied adaptation processes 
using protein regulatory networks and parameter values so that there exists
a large number of attracting states (attractors) of on-off expression patterns.  
The concept that cellular states can be interpreted as attractors has
a long history \cite{Kauffman}, and is supported by several 
experimental studies \cite{Ouyang, Huang1, Huang2}. 
We carried out numerical experiments of the model using several different 
sets of parameter values and choosing thousand of different randomly generated 
reaction networks.
As results, we found that 
the adaptation processes triggered by stochastic noise are generally observed, 
as long as a large number of attractors exists in regulatory dynamics.
In the next section, we present the typical behaviors obtained by using 
networks consisted of $n=96$ proteins and $m=32$ metabolic substrates.

\subsection{Simulation Results}

In Fig.2, an example of such selection process of states is shown by
taking $\sigma=0.2$.  Time series of expression levels of
arbitrarily chosen proteins and growth rate of the cell 
are plotted in Fig.2(a) and (b), respectively.
In the example, 
cells start from a relatively low growth rate.
In such state, stochasticity dominates the time evolution of 
protein levels.
After itinerating among various expression patterns, 
the cellular dynamics finds itself in a state with a 
higher growth rate. Such
transition repeats until the growth rate becomes highly regular.
Once a gene network that provides optimal 
growth is selected, the system maintained it over time.

This selection of higher growth states is observed for all of a thousand networks we simulated.  It also works independently of initial conditions.
As the final state depends on the initial condition, we have
computed the distribution of the final growth rate 
reached from randomly chosen initial conditions.
The distribution of final growth
rate thus obtained is plotted in Fig.3(a).
In the case without noise, i.e., $\sigma=0$, the cellular dynamics
rapidly converge deterministically into an attractor.
In such case, the final growth rates exhibit a broad
distribution as shown in Fig.3(a), representing the wide variety of
the final cellular states.  In contrast, 
under presence of noise ($\sigma=0.2$), the final growth rates exhibit a
relatively sharp distribution, due to the selection process of faster
growth states as we have seen in Fig.2.

Note that once one of the expression patterns is selected as an
attractor, the flux pattern on the metabolic network is uniquely determined. 
As a result, the cellular growth rate $v_g$ is also fixed, which in turn
affects the protein expression dynamics. 
Here the influence of noise depends on the growth rate $v_g$ for each attractor. 
When $v_g$ is small, the deterministic part of
protein expression dynamics (i.e., the first and second terms of r.h.s.
in eq.(1)) is small, so that the stochastic part in the dynamics
is relatively dominant in the protein expression dynamics. 
Then, the probability to escape the attractor due to
fluctuation is large.
In contrast, when the growth rate $v_g$ is large in the attractor, 
the magnitude of the deterministic part of
expression dynamics is larger than that of the stochastic part. As a
result, the probability to escape the state becomes small.
In Fig.3(b), the relationship between the growth rate $v_g$ and the
probability of an escape to an attractor within a period of
time is displayed.  The probability is high when cells are growing slowly and
conversely low when cells are growing rapidly.
It follows naturally from this relationship 
that cells drift with a directional bias toward a higher growth rate.
Hence, as long as the deterministic part of gene expression (i.e., synthesis minus dilution) increases with the growth rate $v_g$ while the noise amplitude has a $v_g$-independent part, the selection of attractors with higher growth rates generally follows\footnote[5]{In \cite{Kashiwagi-Yomo}, selection of the adaptive attractor between bistable states by noise is demonstrated, by introducing phenomenological activity that governs the synthesis and degradation of proteins.}.

The emergence of the selection process as presented in Fig.2 is not
restricted to a specific environmental condition. Instead, 
the mechanism is a general one ensured by 
the physical limitation of the replicating system.
The mechanism makes it inevitable for the cells to seek states with (nearly) optimal growth
independently of environmental context.
To show the adaptation
process over several environmental conditions, we have computed the
temporal evolution of our model, by changing nutrient conditions
i.e., by updating a set of substrates having nonzero $Y_i$, successively.
We have plotted in Fig.4, a time series of protein expressions and the growth
rate, while environmental conditions are changed at the time points indicated by arrows.
After the environmental changes, the fluctuation in expression dynamics
is observed.
This increase in fluctuation continues, until the
cell finally finds a state that ensures a high
growth rate.  Adaptation to a novel environment is thus possible.

Next, we investigate how this noise-driven adaptation depends on 
the noise amplitude. In Fig.5, the
final growth rate $v_g$ is plotted against the noise amplitude
$\sigma$.  For small noise amplitude ($\sigma<10^{-2}$), the final growth
rates are broadly distributed, since cells cannot escape 
from the first attracting state that they encounter.
On the other hand, when the noise amplitude is larger 
($\sigma>1$), the final growth rates again exhibit a
broad range distribution, 
because the cellular state continues to
change without settling into any attractor. In the
intermediate range of the noise strength $10^{-2}<\sigma<1$, 
such cellular states are selected that have significantly higher growth rates than
those found in the other noise ranges. This shift of the final growth rate is 
due to the selection of cellular states by fluctuations, as shown in
Fig.2.

Stability of a given attractor against noise is estimated by
whether the first two terms in eq.(1) are larger than the noise term.
One can roughly estimate that the
stability changes at around $v_g\times O(x) \sim \sigma^2$, where $x$ represent 
the protein expression represented in eq.(1).
If the former term is larger for attractors with higher growth rates,
and smaller for other attractors with lower growth rates, then the
former attractors will be selected.  Considering that the term $O(x)$
is about 0.1 $\sim$ 1, higher growth
rates are selected when $\sigma^2$ exceeds $ \min(v_g)
(0.1\sim1)$, while thus no longer occurs when $\sigma^2> \max(v_g)
(0.1\sim1)$ where all the states are visited randomly
(Here $max$ and $min$ represent the maximum and minimum of $v_g$ over 
attractors, respectively).
The selection
works within the range of noise amplitude $\min(v_g)<\sigma^2/(0.1\sim 1) < \max(v_g)$.
This estimate is consistent with the numerical simulation.

\section{Discussion}

Numerical simulation and analysis of our model demonstrated how
stochastic fluctuations in cellular reaction dynamics result in the
selection of cellular states with higher growth rates.
The selection works for any initial cellular state and environmental conditions, and also works generally irrespectively of choice 
in the model parameters and the topology of reaction network.
For example, if the reaction coefficient of
metabolic reactions changes from $\epsilon = 0.1$ to $10$, the selection of
higher growth states still occurs, although the cells take larger time to approach the
final attracting states.
The selection of an attractor with higher growth rates operates as long as the
cellular states can hop between cellular states by stochastic
fluctuations of protein expressions and there is negative correlation between the cellular
growth rate and time to escape the corresponding states. 
In our model,
the negative correlation is introduced, since
both the synthesis and dilution (degradation) of proteins are proportional to the
growth rate, while the noise amplitude is independent of the growth rate.
Since the dependence of the synthesis and dilution rates on the growth rate is natural, the adaptive attractor selection discussed here is expected to be generally observed for cells that can grow.

With regards to the noise term, the selection mechanism to be discussed 
works in various forms of it, as long as the noise
does not vanish with the growth rate, or in
other words, as long as a certain amplitude of the noise is maintained 
in the non-adaptive state.
For example, we have simulated a model with additional noise term
$\sqrt{v_g} \eta(t)$ in addition to the noise in eq.(1), 
and confirmed that the present adaptive attractor selection still works.

On the other hand, 
if the variance of total noise increased linearly with the growth rate $v_g$, 
the present selection would not work.
When the noise is originated only in the growth-dependent reaction, one might think that this 
is the case.
However, as long as there is basal process for the protein synthesis (and degradation)
even when a cell does not grow, there should exist 
a growth independent part in the noise as in eq.(1).
Although such part of noise has not been measured separately, the 
fact that the synthesis of mRNAs, proteins and metabolites are 
maintained even in the stationary phase of a cell \cite{statonary_phase} suggests that 
there exists a growth-independent part which contributes to the noise.
As long as such growth-independent part exists in the noise, the present mechanism works.

To confirm the generality of the selection mechanism, we have also
simulated a stochastic model by adopting Gillespie algorithm \cite{Gillespie}.  
Due to technical limitation in the computational speed, 
we have simulated a simpler model with few degrees of freedom that allows
for only two attractors in the regulatory dynamics.
As results, we observed that as long as 
the noise does not vanish with the rates of synthesis and degradation of proteins,
higher growth rates are chosen in agreement with 
the simulation of the Langevin equation (eq.(1)). This suggests that 
selection process for a higher growth rate 
works if the number of molecules in a cell is not too large.

The magnitude of protein expression noise quantified by coefficient of variation
could be in the order of $0.1 \sim 0.01$, as shown in Ref.\cite{barkai}.
In some cases it is suggested that the fluctuation is large enough to force cells to back 
and force between discrete states \cite{Oudenaarden}.
This magnitude of noise is within the range of our estimate
required for the attractor selection mechanism, 
although it is necessary to measure the magnitude of the basal noise.
To clarify this point, further experiments about relationship between 
growth rate and fluctuation of protein abundances are necessary.

The results in our study provide an explanation for the establishment of 
the optimal growth rate in the metabolic reaction
networks, to that proposed by Palsson and his colleagues \cite{Palsson1,Palsson2,Palsson3}.
In their model, it is suggested that a metabolic network is
organized so that the growth rate is optimized under given conditions.
For example, it was shown that {\sl E. coli} strains 
with a deletion of a single metabolic gene can adapt to several 
environmental conditions, and that the value of the
final growth rate is consistent with that calculated as an optimal 
growth rate in these perturbed metabolic networks 
and environmental conditions \cite{Palsson3}.
The observed adjustments of metabolic fluxes often occur
within several days, 
suggesting that such adaptation process is not caused 
by selection of mutants having higher division rate
under the given condition.
This means that these bacteria adjust their intra-cellular state
to optimize the growth rate, even though 
they have never experienced such perturbation to the metabolic network.

The result presented in this paper provides a possible
mechanism for selecting a cellular state with optimal growth rate,
for a variety of environmental conditions.
An important point here is that the
presented mechanism requires no 
fine tuning of regulatory
machineries.
As long as the cellular states are perturbed 
sufficiently by the stochastic protein expressions, 
any replicating network will establish 
a negative correlation between
the growth rate and the escape probability from the corresponding
cellular state. Thus we propose that the adaptive attractor selection be at work 
behind the observed regulations of metabolic
fluxes leading to optimal growth rate.

The merit of the present adaptive attractor selection induced by, and
optimizing, growth lies in its generality.  The mechanism
can work without fine-tuning through evolution.
Indeed, it makes adaptation possible to novel environment that the species has not experienced through the course of evolution.  Note that organisms have to survive by adapting to new environment even before specific signal transduction network has developed.  Our mechanism provides such general and non-specific `proto-adaptarion'.

Of course, there are demerits in our mechanism also.  If the difference 
in the growth rates between the two adaptive states is small,
the present mechanism cannot distinguish them.  Either of these
can be selected.  Hence it does not work for very fine selection.
Also, the selection process is expected to be not so fast, as it is stochastic,
compared with the signal transduction mechanism.  Hence, for
the environmental condition that an organism frequently  encounters,
cells have likely developed a sophisticated 
sensory and signal transduction 
network to regulate cellular states to cope with 
changes in environmental conditions.

The growth-induced and fluctuation-based selection of cellular states presented in this
study has not been confirmed in real biological systems so far. To
clarify it, further experiments of regulatory dynamics are
required. There are two possible strategies for such experiments. One
is the construction of the selection of adaptive cellular states by using
artificial gene networks, as demonstrated in Ref.\cite{Kashiwagi-Yomo}. In this
approach, one can introduce a gene network disconnected from the
existing signal transduction networks,
and investigate whether the artificial gene network exhibits selection of
a higher growth state. Another possibility is to study response of cells
with respect to environmental changes that the cells have never faced, 
or response of cells in which known regulatory machineries are destroyed.
In both cases, by investigating the response of cells,
one can examine if cells show adaptive behavior to environmental change,
without the sophisticated regulatory 
machineries, but by utilizing the fluctuation based selection of a higher growth state,
as presented in this paper.

We would like to thank T. Yomo and S. Sawai
for stimulating discussions and 
critical reading of the manuscript.
This research was supported in part by "Special Coordination Funds 
for Promoting Science and Technology: Yuragi Project" of the Ministry 
of Education, Culture, Sports, Science and Technology, Japan.

\newpage

\begin{figure}[tbp]
\begin{center}
\includegraphics[width=9cm,height=9cm]{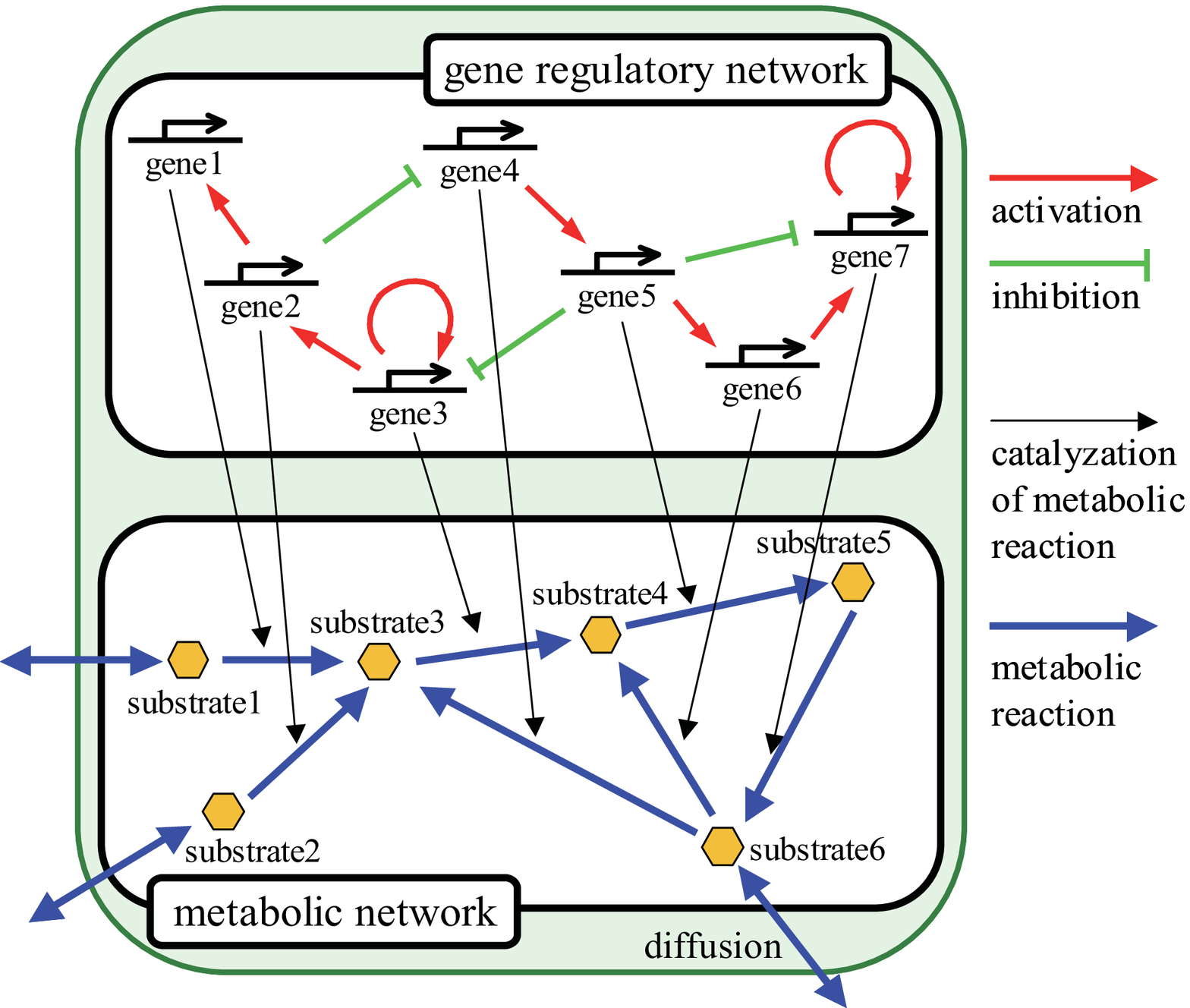}
\caption{
Schematic representation of our cell model.
The model consists of two networks, i.e., 
a gene regulatory network and a metabolic network.
As an example, simple networks consisted of $n=7$ genes and $m=6$ metabolic substrates
are shown.
The red arrows in the regulatory network represent 
activation of expressions, while green lines with  blunt ends 
represent inhibition.
The arrows from a gene to itself mean autoregulation of expressions.
As a result of these regulatory interactions, 
the dynamics of expression levels of proteins have multiple attractors.
The metabolic reactions, represented by blue arrows, are controlled
by expression levels of corresponding proteins.
The correspondence between metabolic reactions and gene products (proteins) are
shown by the black thin arrows.
The regulatory matrix $W_{ij}$ of the presented network takes
$W_{21} = W_{32} = W_{33} = W_{45} = W_{56} = W_{67} = W_{77} = 1$, 
$W_{24} = W_{53} = W_{57} = -1$, and 0 otherwise.
The reaction matrix $Con(i,j,k)$ metabolic network takes a value 1 
for the elements $(1,3,1)(2,3,2)(3,4,3)(6,3,4)(4,5,5)(6,4,6)(5,6,7)$, 
and 0 otherwise. In the present paper, we adopt a much larger network with $n=96$ genes 
and $m=32$ substrates.
}
\end{center}
\end{figure}

\begin{figure}[tbp]
\begin{center}
\includegraphics[width=9cm,height=12cm]{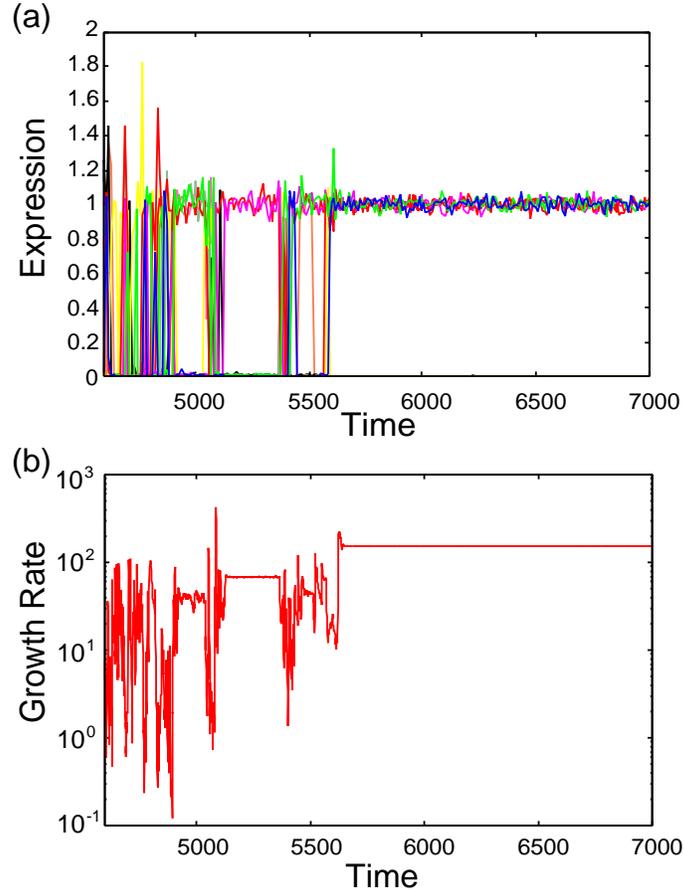}
\caption{
(a): Time series of protein expressions $x_i(t)$.
10 out of 96 protein species are displayed.
The vertical axis represents the expression levels of proteins 
and the horizontal axis represents time.
(b): Change in growth rate $v_g$ observed during the 
time interval shown in (a).
Initially, the growth rate of the cell fluctuates
due to highly stochastic time course of protein expression.
After a few short lived nearly optimal states (c.f. $4800 \sim 5600$ time steps),
the cell finds a state of protein expression that realizes a high rate of growth.
The parameters are $\theta=0.5$, $\mu=10$, $\rho_a=\rho_i=0.03$, $\epsilon=0.1$, 
 and $D=1.0$.
In addition, we enhanced the rate of positive autoregulatory paths,
i.e., $W_{ii}=1$ for i-th gene, 
so that the regulatory network has multiple attractors.
In the simulations, $30\%$ of activating paths are chosen as 
autoragulatory paths.
}
\end{center}
\end{figure}

\begin{figure}[tbp]
\begin{center}
\includegraphics[width=9cm,height=12cm]{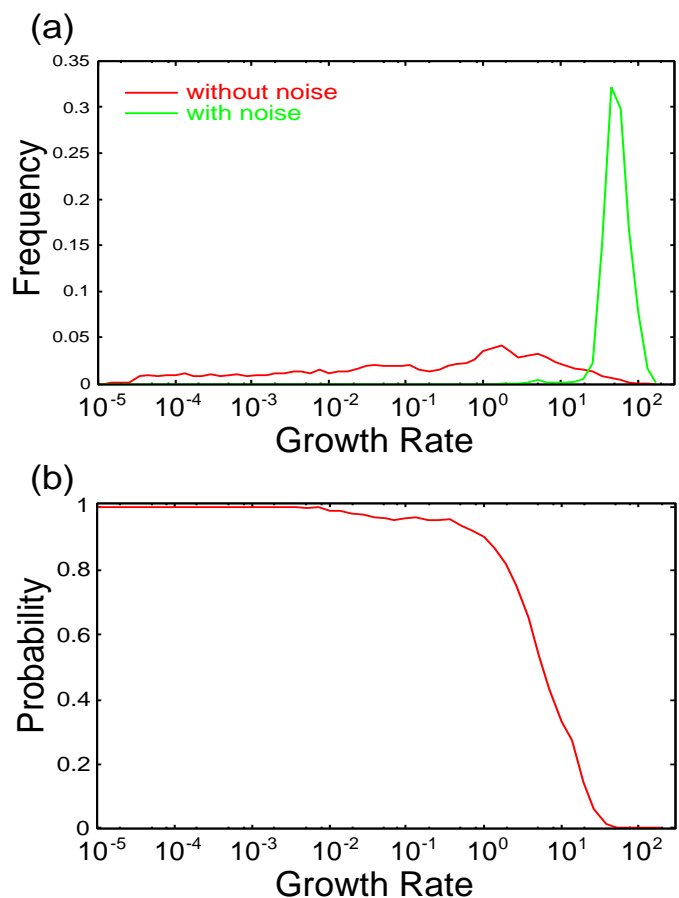}
\caption{
(a): The distribution of growth rate.
Starting from randomly chosen $10^5$ initial conditions, 
the distribution of growth rates after $10^5$ time steps are computed 
with and without noise ($\sigma=0.2$).
(b):
Relationship between the growth rate $v_g$ and 
the probability to escape an attractor
with a certain period of time.
The probability is computed by $10^5$ trials starting 
from randomly chosen initial conditions.
After a cell reaches a stable state, 
noise ($\sigma = 0.2$) is added and the time 
it takes the cell to escape from 
the corresponding attractor is measured.
The y-axis represents the probability 
that the cellular state is kicked out of the original state
within $10^3$ time steps, and 
the horizontal axis shows the growth rate $v_g$ of the original state.
}
\end{center}
\end{figure}

\begin{figure}[tbp]
\begin{center}
\includegraphics[width=9cm,height=12cm]{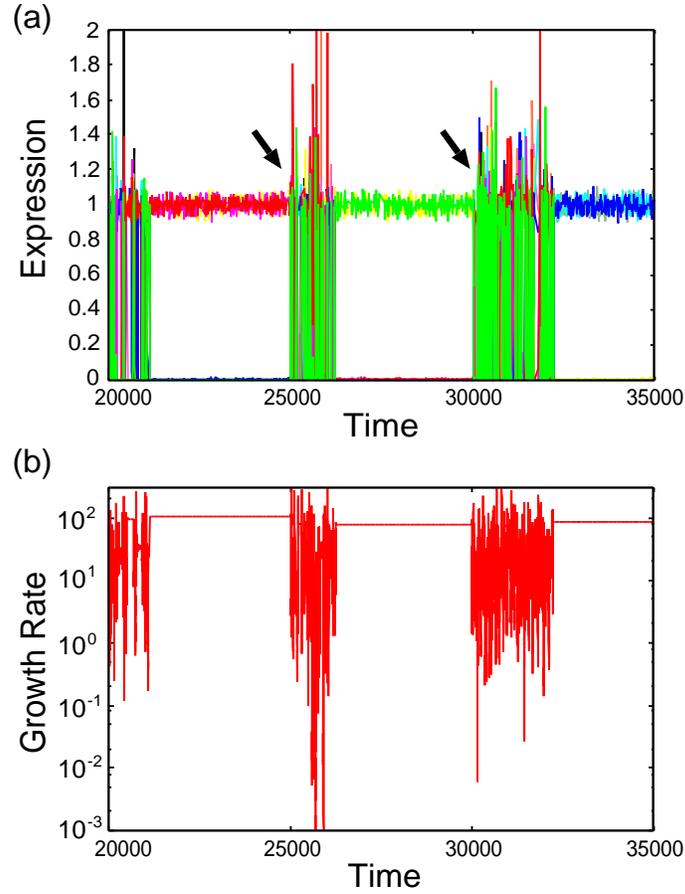}
\caption{
(a): Time series of protein expressions $x_i(t)$ 
when the environmental condition is altered.
The environmental conditions, i.e., substrates having nonzero $Y_i$, 
are changed at time points indicated by arrows.
(b): Change of growth rate $v_g$ in the same time interval as (a).
After the environmental changes, both protein expression levels and the growth rate 
fluctuate until the cell finds a state of protein expression that realizes a high growth rate.
In the simulation, the noise amplitude $\sigma = 0.2$.
}
\end{center}
\end{figure}

\begin{figure}[tbp]
\begin{center}
\includegraphics[width=9cm,height=7cm]{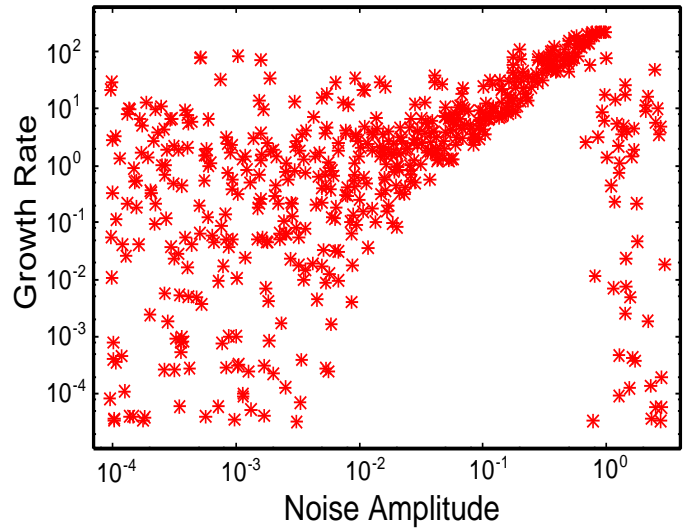}
\caption{
The relationship between the noise amplitude $\sigma$ and the growth rate $v_g$.
Starting from randomly chosen initial conditions 
against the noise amplitude $\sigma$ ranging 
$10^{-4} < \sigma < 3\times10^0$, the growth rates $v_g$ after $10^5$ time steps are plotted.
In the intermediate range of the noise strength $10^{-2}<\sigma<1$, 
cellular states with high growth rates are selected among a huge number of 
possible cellular states, as depicted in Fig.2.
}
\end{center}
\end{figure}

\end{document}